\documentclass{jaa}
\pdfoutput=1 
\usepackage{lastpage}
\usepackage{graphicx}	
\usepackage{amsmath}	
\usepackage{amssymb}	
\usepackage{adjustbox}
\usepackage{longtable}
\usepackage{multirow}
\usepackage{float}
\usepackage{natbib}
\usepackage{hyperref} 
\def\apjs{{\it Astrophys.~J.~Suppl.}}
\def\apj{{\it Astrophys.~J.}}
\def\aj{{\it Astronom.~J.}}
\def\apjl{{\it Astrophys.~J.~Lett.}}

\def\mnras{{\it Mon.~Not.~R.~Astron.~Soc.}}

\def\aap{{\it Astron.~Astrophys.}}

\begin{document}\sloppy

\title{
Study of Classical Be stars in open clusters older than 100 Myr}


\author{Madhu Kashyap Jagadeesh\textsuperscript{1}, Blesson Mathew\textsuperscript{1}, KT Paul \textsuperscript{1,*}, Gourav Banerjee \textsuperscript{1}, Annapurni Subramaniam \textsuperscript{2}, and
R. Arun \textsuperscript{1}}
\affilOne{\textsuperscript{1}Department of Physics and Electronics, Christ (Deemed to be University), Bengaluru 560 029, India.\\}
\affilTwo{\textsuperscript{2}Indian Institute of Astrophysics, Bengaluru 560034, India.}


\twocolumn[{

\maketitle

\corres{paul.kt@christuniversity.in}


\begin{abstract}
We performed the slitless spectroscopic survey to identify Classical (CBe) stars in open clusters older than 100 Myr.
 Observing a sample of 71 open clusters, we identified 13 CBe stars in 11 open clusters, one of, which (TYC 2679-432-1) is a new detection. The 13 CBe stars show both H$\alpha$ in emission and IR excess, which confirm that they possess gaseous circumstellar discs. Two more CBe stars are found to exhibit H$\alpha$ in absorption for the first time, indicating that might be passing through disc-less episode presently. The spectral type estimation is done for all these 15 stars and it is noticed that they belong to B0.5 -- B8 type. Moreover, we found that the distribution of our sample is peaking near late B-types as expected.
\end{abstract}


}]


\doinum{12.3456/s78910-011-012-3}
\artcitid{\#\#\#\#}
\volnum{000}
\year{0000}
\pgrange{1--}
\setcounter{page}{1}
\lp{1}

\section{Introduction}
\cite{1987Collins} defined a Classical Be (CBe hereafter) star as ‘a non-supergiant B-type star whose spectrum has, or had at some time, one or more Balmer lines in emission'. These are main sequence stars surrounded by a geometrically thin, gaseous, equatorial, decretion disc that orbits the central star in Keplerian rotation (\citealt{2007Meilland}), belonging to luminosity classes III-V. The first viable model of CBe stars was proposed by \cite{1931Struve}, who suggested that the high rotation rate of such stars result in the ejection of stellar material, thus forming a circumstellar disc. A comprehensive review of CBe star studies can be found in \cite{2013Rivinius} and \cite{2003Porter}.

CBe star discs, unlike protoplanetary dusty envelopes that orbit young stars, are not shrouded in dust. Hence, CBe stars provide a natural laboratory to study circumstellar discs. However, the disc formation mechanism in CBe stars - known as the ‘Be phenomenon’, remains as an unsolved problem. If the mass is ejected from the star (due to the combined effect of rotation, non-radial pulsations etc.), it can be in orbit around the star and form a circumstellar disc. This is well explained by the viscous decretion disc (VDD) model proposed by \cite{1991Lee} and confirmed by \cite{2012Carciofi}.

Appearance of emission lines of various elements (such as hydrogen, iron, oxygen, helium, calcium, etc) is a common observable property of CBe stars. Spectral analysis of these lines provide a wealth of information about the geometry and kinematics of the gaseous disc and several properties of the central star itself. As a result, numerous spectroscopic surveys have been carried out, both in the optical (e.g. \citealt{1982AndrillatF, 1986Hanuschik, 1986Dachs, 2012Koubsky, 2017Arcos, 2019Klement}) and in the near-infrared (e.g. \citealt{2000Clark, 2001Steele, 2011Granada}) bands to characterize CBe star discs and to better understand the ‘Be phenomenon'. Focusing on the same aspects, several other spectroscopic surveys have also been performed for CBe stars located in different environments, such as fields (\citealt{2020Gourav}), clusters (\citealt{2008Mathew}) and extra-galactic regime like the Magellanic clouds (\citealt{2012Paul}).
Open clusters contain stars of similar age, thus serving as nice test-beds to study CBe stars and their evolutionary stages. Few prominent works on studying CBe stars in open clusters are by McSwain \& Gies (2005), \cite{2008Mathew} and Martayan et al. (2010). \cite{2008Mathew} identified 152 CBe stars in 42 open clusters younger than 100 Myr using slitless spectroscopy technique. The spectral features observed in these stars were presented in \cite{2011Mathew}. This study motivated us to perform a similar kind of spectroscopic survey to identify CBe stars in open clusters older than 100 Myr as this will include the sample of late B-type stars.

\begin{figure*}[h!]
\centering 
\includegraphics[width=0.5\columnwidth]{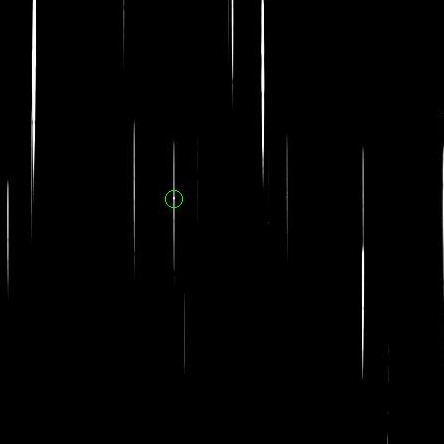}
\includegraphics[width=0.5\columnwidth]{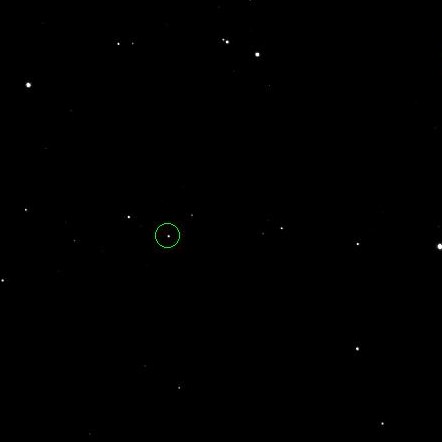}
\includegraphics[width=0.65\columnwidth]{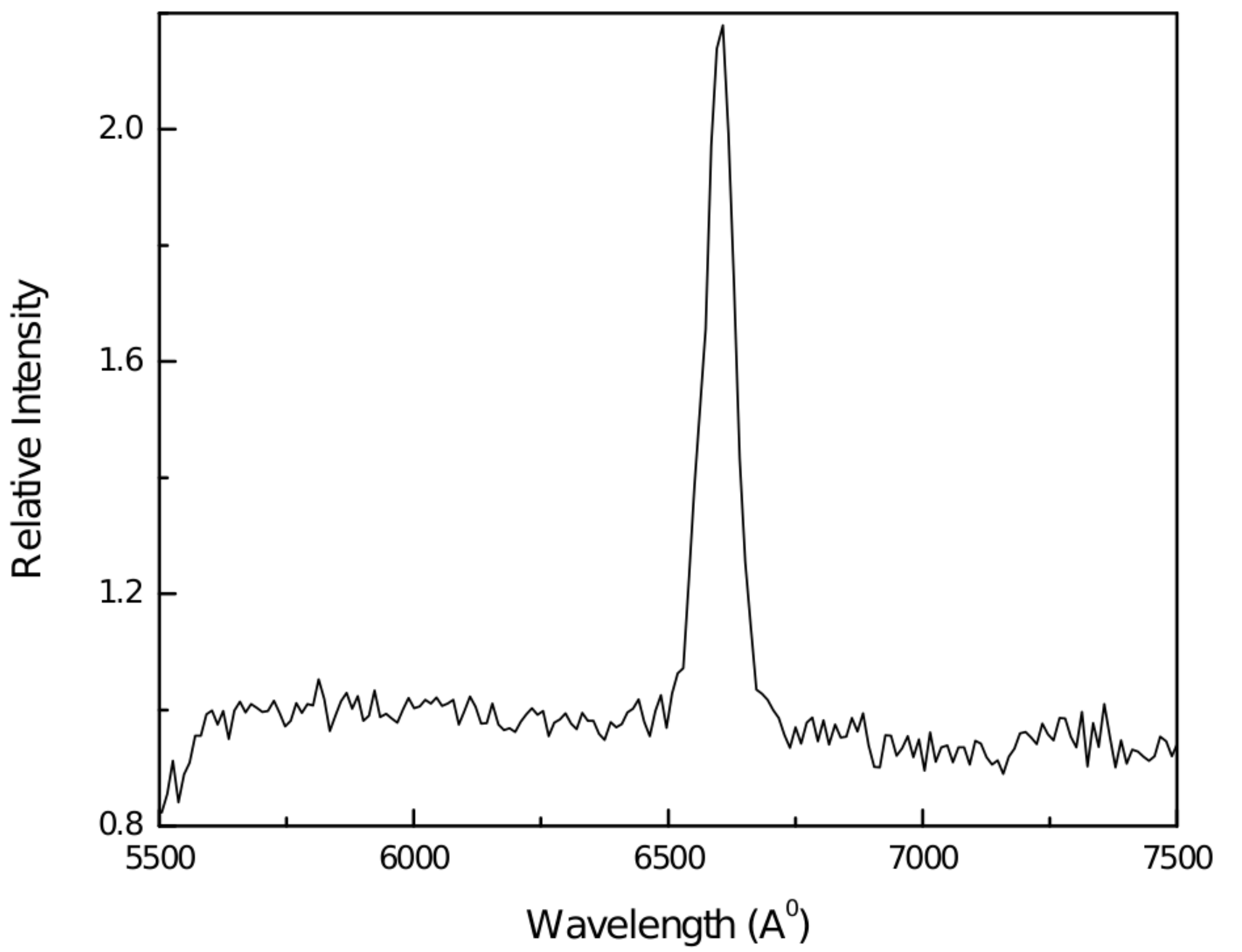}
\caption{A sample dispersed image of the cluster Berkeley 47 is presented in the left panel. The middle panel is the corresponding R-band image of the same cluster, while the right panel shows the slitless spectrum of the star TYC 1605-346-1 exhibiting H$\alpha$ in emission} 
\end{figure*}

In the present study, we performed slitless spectroscopy of 71 open clusters older than 100 Myr and identified 13 CBe stars in 11 clusters. According to our knowledge, this is the first study to detect and characterize CBe stars in clusters older than 100 Myr. Out of our 13 identified CBe stars, one star (TYC 2679-432-1) is a new detection. Moreover, we also found two CBe stars showing H$\alpha$ in absorption for the first time. The paper is organised as follows: Sect. 2 provides the observation techniques used for our study. In Sect. 3, we describe the major results and analysis done in this work. Prominent results of the study are summarised in Sect. 4.

\section{Observations}
\label{sect:Obs}
Spectroscopic observations of our program clusters were performed using the HFOSC instrument mounted on the 2.1-m Himalayan Chandra Telescope (HCT) situated at Hanle, Ladakh, India. HCT is operated by the Indian Institute of Astrophysics, Bangalore\footnote{$(http://www.iiap.res.in/)$}. We observed 71 clusters in the Milky Way galaxy during July 2015 to April 2017 using slitless spectroscopy technique. This technique was previously used in a survey of young open clusters by \cite{2008Mathew}. The log of observations is presented in Table 1.

The CCD used for the present study contains $2K \times 4K$ pixels, of which the central $2K \times 2K$ pixels were used for imaging. The pixel size is $15\mu$m with a plate scale of 0.297 arcsec pixel$^{-1}$. The total observed area is $10 \times 10$ arcmin$^2$, approximately. The R band filter $(7100 \AA, BW=2200 \AA)$ and Grism 5 $(5200--10300 \AA)$ were used in combination to obtain the slitless spectrum. Data obtained were reduced using standard tasks in the Image Reduction and Analysis Facility (IRAF) software. A sample slitless spectrum of the star TYC 1605-346-1 exhibiting H$\alpha$ in emission is shown in Fig. 1. The corresponding dispersed (R + Grism 5) and R-band images of its host cluster Berkeley 47 is also shown in the same figure.

\section{Results and Discussion}
We surveyed the literature to check whether CBe stars have been identified previously in clusters above 100 Myr or not? It is found that 16 CBe stars, other than ours, have been detected in 6 clusters. Table 2 lists the names of these 16 stars along with their host clusters and the corresponding literature, which identified them. Now, our identification of another 13 increases the sample of known CBe stars in clusters older than 100 Myr to 29.

In Sect. 3.1, we present the details of each of our identified 13 CBe stars having age above 100 Myr. Table 3 presents the list of these 13 stars in 11 clusters along with their coordinates and V magnitudes, as obtained from the SIMBAD database. In Sect. 3.2 we present the details of two stars showing H$\alpha$ in absorption and compared it with previous studies where they were listed as Be stars. Estimation of spectral types of our identified stars is performed in Sect. 3.3. The infrared excess (hereafter IR excess) study, color-magnitude diagram and spectral type distribution of our identified sample are presented in Sect. 3.4, 3.5 and 3.6, respectively.

\subsection{CBe stars identified from the present study}
\subsubsection{SS 16}:
Hosted by the cluster Tombaugh 5 (\citealt{2011Zdanavi}), SS 16 was detected as a star showing H$\alpha$ in emission by \cite{1977Stephenson}. Although less studied, it was later listed in the studies of H$\alpha$ emission stars by \cite{1983Coyne}, \cite{1997Kohoutek, 1999Kohoutek}. \cite{2011Zdanavi} estimated its spectral type to be B2. The distance, age and E(B-V) values for its parent cluster Tombaugh 5 are 1750 pc, 199 Myr and 0.8 mag (\citealt{2004Lata}), respectively.

\subsubsection{HD 280460}:
The MWC bright star catalog detected HD 280460 as a B-type star with bright hydrogen lines (\citealt{1949Merrill}). \cite{1997Kohoutek, 1999Kohoutek} also observed this star exhibiting H$\alpha$ in emission. \cite{1995Nesterov} reported its spectral type as B8. HD 280460 is hosted by the cluster NGC 1778, which is situated at a distance of 1062 pc (\citealt{1961Hoag, 1975Joshi}). The corresponding age and E(B-V) values of the cluster are 158 Myr and 0.34 mag as reported by \cite{1961Hoag}. Subsequently, \cite{2018Cantat} confirmed the membership of HD 280460 in NGC 1778. This star is also listed in the studies of \cite{2015Paunzen}. 

\subsubsection{HD 280461}:
HD 280461 is a less studied CBe star listed in the studies of \cite{1982Mermilliod} and is a member of the open cluster NGC 1778 (\citealt{1961Hoag}). Its spectral type was estimated as B6 by \cite{1961Hoag}. This star is also listed in the studies of \cite{1995Ahumada}, \cite{2009Skiff}, \cite{2011Guerrero} and \cite{2015Paunzen}.

\subsubsection{HD 280462}:
HD 280462 was cataloged as a multiple trapezium type system by \cite{1978Salukvadze} and \cite{1979Salukvadze}. Later, it was listed in the CBe star studies of \cite{1982Mermilliod}. \cite{1995Nesterov} estimated its spectral type to be B9. 
Located in the cluster NGC 1778, the cluster membership of HD 280462 was established by \cite{2018Cantat}.

\subsubsection{LS V +24 11}:
\cite{1999Kohoutek} detected LS V +24 11 to show H$\alpha$ in emission.
Situated in the cluster IC 2156, its distance, age and E(B-V) values are 2100 pc, 251 Myr and 0.67 mag (\citealt{2009Tadross}), respectively. \cite{2003Reed} has identified LS V +24 11 as an OB star, whereas \cite{2015Armstrong} also listed it. Our study further confirms it as a CBe star.

\subsubsection{SS 398}:
Detected by \cite{1977Stephenson} as a H$\alpha$ emission star, SS 398 is almost unstudied. It is located in the cluster Ruprecht 144, which is situated at a distance of 1727 pc and possesses age and E(B-V) values of 158 Myr and 0.70 mag (\citealt{2019Clari}), respectively.

\subsubsection{GSC 5692-0543}:
GSC 5692-0543 was detected by \cite{2005McSwain} to be a CBe star through photometric analysis. Hosted by the Trumpler 34 cluster, it is located 1738 pc away from us (\citealt{2005McSwain}). The corresponding age and E(B-V) values for Trumpler 34 are 125 Myr and 0.56 mag, respectively (\citealt{2005McSwain}).

\subsubsection{BD+10 3698}:
Detected to be an emission-line star by \cite{1997Kohoutek} and also listed in the studies of \cite{1999Kohoutek}, BD+10 3698 is located in the cluster NGC 6709. The host cluster NGC 6709 is located at a distance of 1075 pc. Its age and E(B-V) values are 158 Myr and 0.304 mag (\citealt{1999Subramaniam}). In a subsequent study, \cite{2008Mermilliod} used radial velocity measurement method to confirm the membership of BD+10 3698 in NGC 6709. Later, \cite{2018Cantat} also established the membership of this star in the same cluster.

\subsubsection{TYC 1605-346-1}:
\cite{1997Kohoutek} first identified TYC 1605-346-1 as a star showing H$\alpha$ in emission. Subsequently, \cite{1999Kohoutek} also listed it in the studies of H$\alpha$ emission stars. Located in the cluster Berkeley 47 (\citealt{2010Krone, 2018Cantat}), this is a less studied star. 
Its parent cluster, Berkeley 47 is situated at a distance of 1420 pc, has
 an age of 158 Myr and E(B-V) value of 1.06 mag (\citealt{2010Subramaniam}).

\subsubsection{UCAC3 274-184438}:
UCAC3 274-184438 is hosted by the Berkeley 90 cluster (\citealt{2008Tadross}), which is located at a distance of 2430 pc and has age of 100 Myr, E(B-V) value of 1.15 mag (\citealt{2008Tadross}), respectively. \cite{2011Mathew} first detected it to be a CBe star of spectral type B0. Later, \cite{2018Dimitrov} studied the H$\alpha$ variability of this star and found that the emission profiles appear almost constant with time or are highly variable in intensity.

\subsubsection{LS III +47 37}:
LS III +47 37 is a visual binary situated in the cluster NGC 7067 (\citealt{2017Mongui}). NGC 7067 is located at a distance of 3600 pc, age = 100 Myr and its E(B-V) value is 0.75 mag (\citealt{1965Becker}). LS III +47 37, with Gaia ID 2164725643889832960 is one of the component of this multiple star system, which was detected to be a CBe star with spectral type close to B0.5 (\citealt{2017Mongui}). This star clearly showed H$\alpha$ in emission when we observed it on September 07, 2016.

\subsubsection{GGR 148}:
\cite{1956Gonz} first detected GGR 148 to exhibit H$\alpha$ in emission. Later, it was also detected to be an emission-line star by \cite{1997Kohoutek, 1999Kohoutek}. This star is also listed in the catalog of wide binary and multiple systems of bright stars (\citealt{2019Jim}). 
GGR 148 is hosted by King 20 cluster, which is at a distance of 1900 pc and possesses age and E(B-V) values of 199 Myr and 0.65 mag (\citealt{2010Glushkova}), respectively.

\subsubsection{TYC 2679-432-1}:
This is an unstudied star with only the coordinates, parallax (mas) and flux values in 6 bands (B, V, G, J, H and K) listed in the SIMBAD database. It is hosted by the Berkeley 50 cluster, which is situated at a distance of 2100 pc and has age = 251 Myr and E(B-V) = 0.97 mag (\citealt{2008Tadross}).

Through our present survey, we identified TYC 2679-432-1 to be a CBe star for the first time. H$\alpha$ was present in emission when we observed it on September 07, 2016. 

\subsection{CBe stars showing H$\alpha$ in absorption}
CBe stars are known to show variability in spectral line profiles. The extreme case of such a variability is the disappearance of H$\alpha$ emission line, indicative of disc-less state in CBe stars. The spectrum will then look like that of a B-type star with photospheric absorption lines.A well studied example is the disc-less episode of X Persei (\citealt{1991Norton, 2013Mathew}). Observations during such a disc-less state can be used to estimate the stellar parameters such as spectral type and luminosity since the spectral lines are unaffected by veiling from the disc (\citealt{1992Fabregat}).

During our survey, we detected 2 stars (apart from identifying 13 CBe stars) where H$\alpha$ is present in absorption. These 2 stars, [KW97] 35-12 and HD 16080, are located in the clusters NGC 6709 and Trumpler 2, respectively. Both of them were found to have shown H$\alpha$ emission in the past. Table 3 presents the coordinates and V magnitudes of these 2 stars.

\subsubsection{[KW97] 35-12}:
A less studied star, it is hosted by the cluster NGC 6709 and was detected by \cite{1997Kohoutek, 1999Kohoutek} to show H$\alpha$ in emission. We observed [KW97] 35-12 on July 08, 2015 and found H$\alpha$ to be present in absorption.

\subsubsection{HD 16080}:
\cite{1965Hoag} estimated the spectral type of HD 16080 to be B7, belonging to the open cluster Trumpler 2. This cluster is at a distance of 651 pc, age is 147 Myr and its E(B-V) value is 0.324 mag (\citealt{1965Hoag}). \cite{1976Schild} and \cite{1978Schild} reported HD 16080 to be a CBe star while studying CBe stars in clusters. Later, it was listed in the Be star studies of \cite{1982Mermilliod}. Subsequently, \cite{1999Kohoutek, 1997Kohoutek} noticed this star to exhibit H$\alpha$ in emission. However, H$\alpha$ was present in absorption when we observed HD 16080 on January 02, 2016. These 2 stars, [KW97] 35-12 and HD 16080, are located in the clusters NGC 6709 and Trumpler 2, respectively. 

Apart from detecting these 2 stars, we also noticed H$\alpha$ in absorption in 5 other stars situated in the cluster NGC 7296. This cluster is located at a distance of 2930 pc and has age of 100 Myr (\citealt{2005Netopil}). These 5 stars of NGC 7296 are classified as $Ae/Be$ stars by \cite{2005Netopil} through photometric study. Surprisingly, all these 5 showed H$\alpha$ in absorption when we observed them. 

We re-estimated the spectral types for all these 7 (2 previously mentioned and the current 5) stars using photometric technique. It was found that the previous 2 ([KW97] 35-12 and HD 16080) and 4 among the rest 5 stars located in NGC 7296 belong to B-type. Hence, our result suggests that [KW97] 35-12 and HD 16080 might be CBe stars currently passing through disc-less state. However, the nature of all these 7 stars can be confirmed through spectroscopic studies in future, which we aim to perform as a follow-up work. 

\subsection{Estimating the spectral type of our identified sample}
Spectra of all the 15 stars (13 identified CBe + 2 stars showing H$\alpha$ in absorption) were obtained in the H$\alpha$ region. So we could not estimate the spectral types for our stars using spectroscopy since the He{\sc i} lines near the blue region are not present. However, literature review shows that estimating the spectral types for CBe stars using photometric technique has been performed widely by various authors such as \cite{1993Halbedel}, \cite{2000Coe}, \cite{2001Torrej}, \cite{2001Beaulieu}, \cite{2005Kahabka}, \cite{2006Blay}, \cite{2014Scott}. Hence, we also estimated the spectral types for all 15 stars using photometric technique and found that they belong within B0.5 -- B8 type (shown in Table 3). For 6 among these 15 stars, spectral type estimation is performed for the first time. For the rest 9 cases, we found previous literature where spectral type estimation was done. 

To estimate the spectral types, we first used the following distance modulus relation to calculate the absolute magnitude (M$_{V}$) of individual stars:

\begin{equation*}
   m_{V} - M_{V} = 5 \log_{10} d - 5 + A_{V}
\end{equation*}

where m$_{V}$ and M$_{V}$ are the apparent and absolute magnitudes of a star, d is its distance in parsec and $A_{V}$ is the interstellar extinction associated with that star. The distance (d) of all our stars are taken from the literature as presented in Sect. 3.1. We obtained their corresponding m$_{V}$ from the Tycho 2 (\citealt{2000H}) and NOMAD (\citealt{2005Zacharias}) catalogs. To find their associated $A_{V}$, we used the following relation:

\begin{equation*}
  A_{V}/E(B-V) = R_{V}
\end{equation*}

where E(B-V) is the color excess whose values are adopted from the literature. $R_{V}$ is the ratio of total-to-selective extinction whose value is adopted as 3.1. The estimated M$_{V}$ values using this method is then matched with the look up table given in \cite{2013Pecaut} to determine the spectral types for our identified stars. 

\subsection{Near-IR Color-Color diagram of our identified CBe stars}
As a next step after estimating the spectral types of our sample stars, we wanted to confirm whether they really possess a circumstellar, gaseous disc or not. In CBe stars, the IR excess is commonly attributed to thermal bremsstrahlung emission from free electrons in a hot, dense, ionized circumstellar disc (\citealt{1974Gehrz,1977Hartmann}). In case of PMS star discs, such as in HAeBe stars, dust is also present (\citealt{1993Gorti, 1998WatersWaelkens}, \citealt{2019Arun}). As a result, the IR excess emission in case of HAeBe stars originate due to the re-radiation from the dust in the disc (\citealt{1992Hillenbrand}). Since dust emission is more intense than thermal bremsstrahlung, IR excess in HAeBe stars are relatively higher than CBe stars. \cite{1984Finkenzeller} suggested that the (H-K$_{S}$) colors of HAeBe stars are found to be greater than 0.4 mag.

We plotted the near-IR color-color diagram (CCDm) for 13 of our identified CBe stars, which is presented in Fig. 2. The J, H, K$_{S}$ magnitudes for our sample are obtained from the 2MASS point source catalog (\citealt{2003Cutri}) using VizieR for a search radius of 3 arcsec. These adopted J, H, K$_{S}$ magnitudes need to be corrected for extinction to represent them in the 2MASS CCDm. The extinction values are estimated using the standard relations given in \cite{1988Bessell}.

In Fig. 2, the red triangles represent 13 of our CBe stars. Similarly, the black dots are CBe stars located within young open clusters below 100 Myr, taken from \cite{2008Mathew}. The solid line represents the main sequence and the dotted line represents the reddening vector, taken from \cite{1983Koornneef}. They are converted to the 2MASS system using the transformation relations from \cite{2001Carpenter}.

\begin{figure}
\centering
\includegraphics[width=7.5cm]{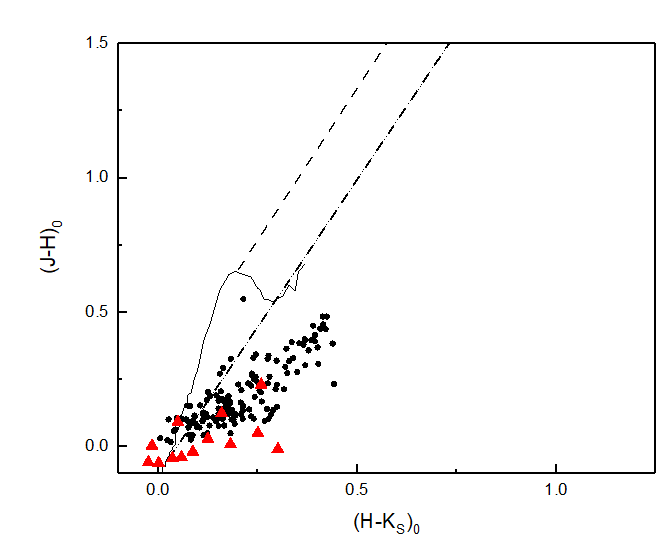} 
\caption{Near-IR color-color diagram of 13 identified CBe stars from our survey. Here, the red triangles represent 13 of our CBe stars, where as the black dots are CBe stars located in} young open clusters below 100 Myr, taken from \cite{2008Mathew}. The solid line in the figure represents the main sequence boundary and the dotted line represents the reddening vector, for, which the established controversial slope of 1.7 is taken from \cite{1983Koornneef}. The data transformation for 2MASS system plotting is performed using \cite{2001Carpenter}.
\end{figure}

It is observed from the figure that our sample of stars are situated in a similar location as that of the previously studied cluster CBe stars by \cite{2008Mathew}. We noticed that they are also exhibiting considerable amount of IR excess, which is a distinctive property of CBe stars. Thus, it is found that all our identified 13 stars show both H$\alpha$ in emission and also IR excess. Hence, this result suggests that these 13 stars possess gaseous, circumstellar discs.

Next, we constructed the Spectral Energy Distribution (SED) plots for all 13 stars to quantify the IR excess. The SEDs are constructed using the U, B, V, J, H, K, Gaia (G, $G_{BP}, G_{RP}$) and WISE bands (W1, W2, W3, W4), which were computed using the VOSA tools (\citealt{2008Bayo}). The SED of one of our sample stars, TYC 2679-432-1, is presented in Fig. 3. It is noticed that the star is showing IR excess in the WISE band region, thus indicating the presence of a gaseous, circumstellar disc surrounding the central star.

\begin{figure}
\centering
\includegraphics[width=1.0\columnwidth]{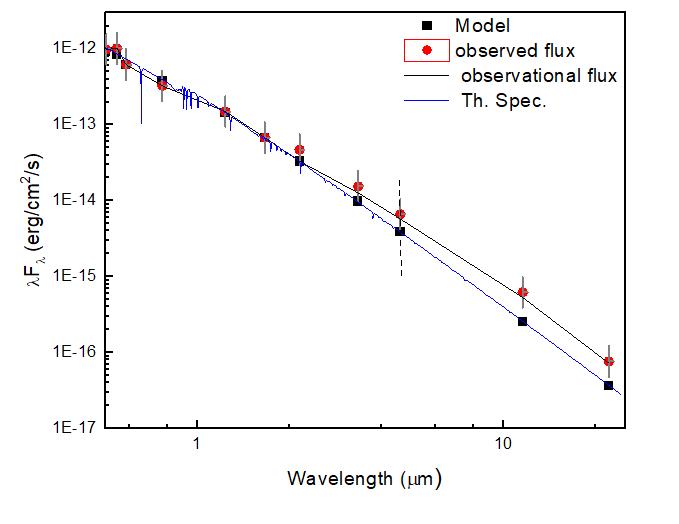}
\caption{SED of our sample CBe star TYC 2679-432-1. For this star, we used the BT-NextGen-AGSS2009 stellar atmospheric model and effective temperature (T$_{eff}$) = 11200K is adopted. It is observed that TYC 2679-432-1 is showing IR excess in the mid IR range indicating the presence of a gaseous circumstellar envelope around it.} 
\end{figure}

\subsection{Analysis using Color-Magnitude diagrams}
Color-Magnitude diagrams (CMDs) are necessary to provide an understanding about the relationship between the evolutionary phase of CBe stars with respect to their age. The CMDs are plotted only for the clusters NGC 6709, NGC 1778 and King 20, for which optical photometry data is available.

The photometric data for NGC 6709 open cluster is obtained from \cite{1999Subramaniam}. The CMD is shown in Fig. 4 with the CBe star and the transient Be star in separate colors. The black filled circles represent the cluster members, the solid line is the main sequence and dotted line represent the MIST MESA (\citealt{2016Choi}; \citealt{2016Dotter}) isochrone with a fitting log(age) of 8.178 (\citealt{1999Subramaniam}). Both the stars are positioned near the turn-off point of the main sequence.

The photometric data for NGC 1778 open cluster is obtained from \cite{2017Sampedro}. After converting the magnitude and color in absolute scale, they are represented in the CMD in Fig. 5. The cluster members are shown in filled circles and the CBe stars HD 280461, HD 280460 and HD 280462 are shown in the CMD in different colors. An isochrone corresponding to a log(age) of 8.155 is shown, which represents the age of the cluster. It is seen that the CBe stars are close to the turn-off point of the cluster.

The photometric data for King 20 open cluster is obtained from \cite{2010Glushkova}. The CMD is shown in Fig. 6 with the cluster members in filled circles and the CBe star GGR 148 shown in red. A ZAMS and an isochrone of log(age) = 8.30, corresponding to the age of the cluster is represented. The CBe star seems to have evolved from the main sequence, although a confirmation from a follow-up spectroscopy is warranted.

From the CMD analysis of the three clusters it is seen that the CBe stars are close to the turn-off region and hence may be evolving off the main sequence. However, in order to provide a proper luminosity classification slit spectrum of these stars is required. Also, it may be noted that optical photometric data needs to be obtained for other clusters to gain evolutionary information of other CBe stars.

\begin{figure}
\centering
\includegraphics[width=1.0\columnwidth]{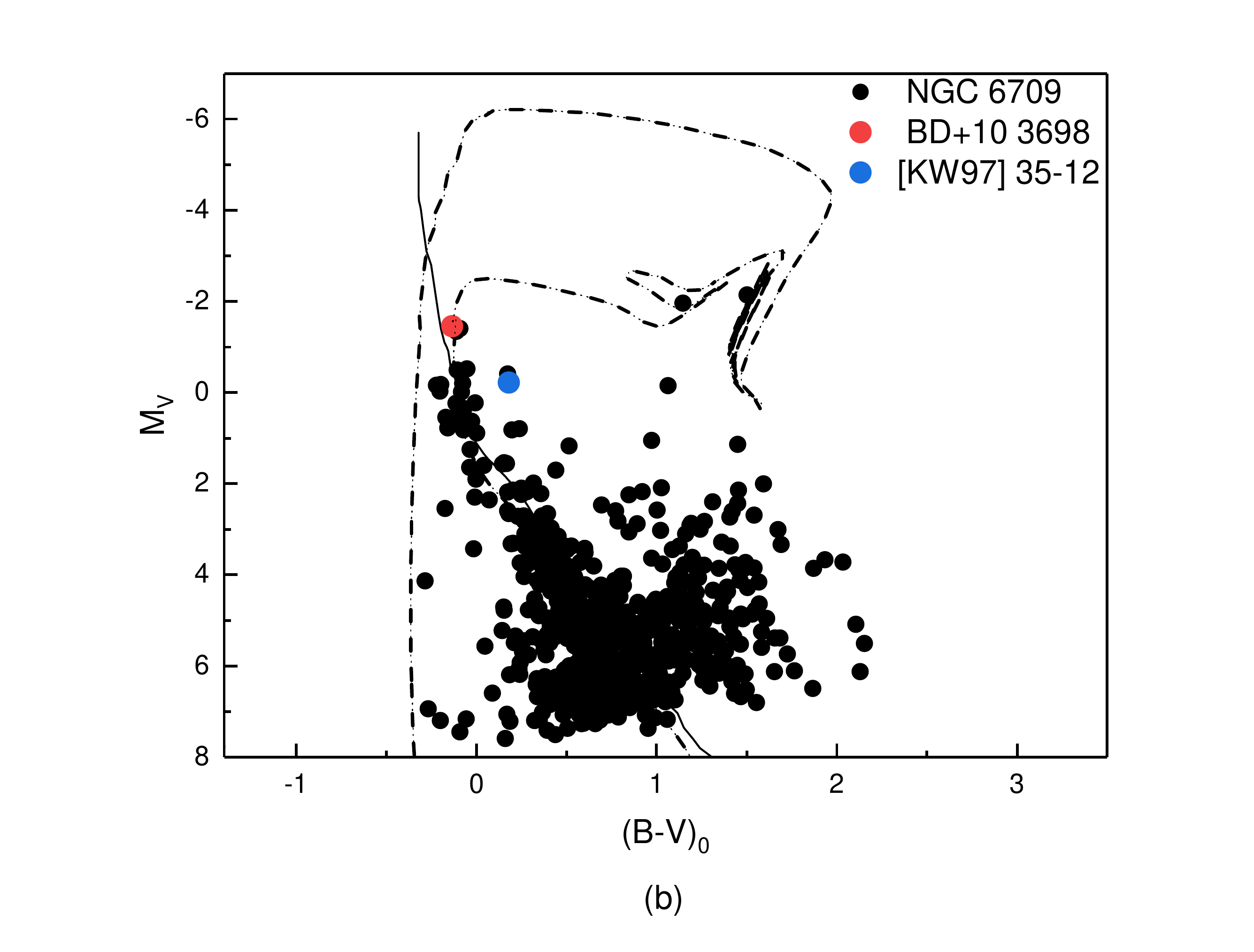}
\caption{CMD for the NGC 6709 open cluster} 
\end{figure}

\begin{figure}
\centering
\includegraphics[width=1.0\columnwidth]{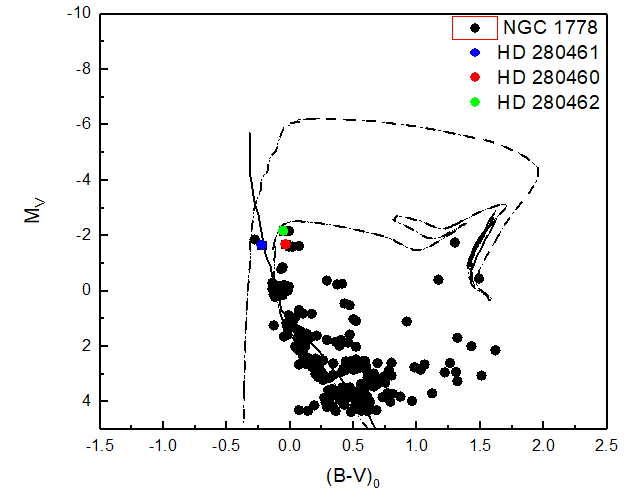}
\caption{CMD for the NGC 1778 open cluster} 
\end{figure}

\begin{figure}
\centering
\includegraphics[width=1.0\columnwidth]{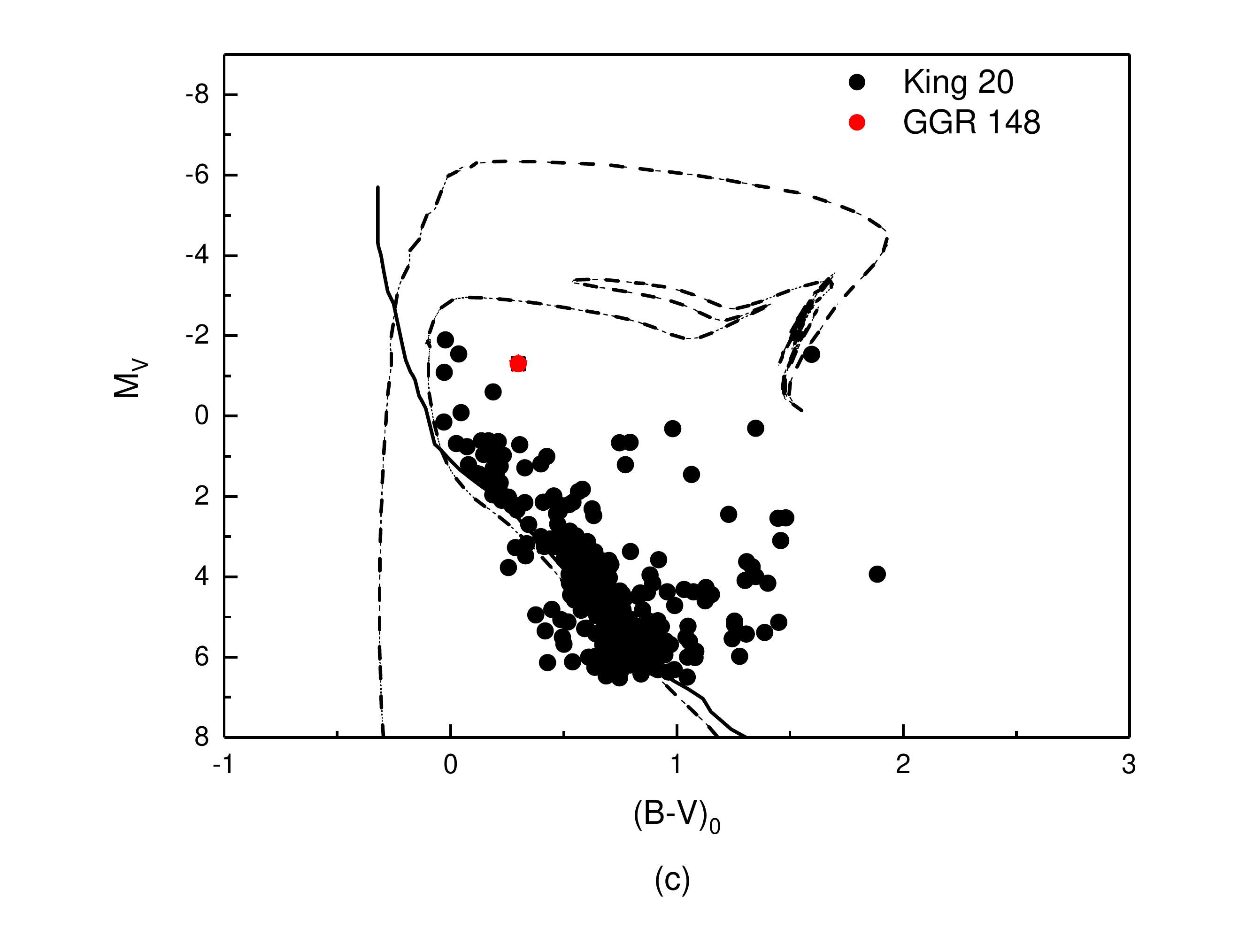}
\caption{CMD for the King 20 open cluster} 
\end{figure}

\subsection{Spectral type distribution of the CBe stars}
Observational studies have shown that CBe stars include a spectral range from O9 to A3. \cite{1982Slettebak} noticed that the spectral type distribution of a sample of 183 CBe stars peaked at B2 spectral type. \cite{1982Mermilliod} reported the distribution of 94 CBe stars in 34 open clusters peaking at spectral types B1–B2 and B7–B8. A similar kind of bimodal distribution, peaking at early and late B-type spectral regions, was noticed by Mathew et al. (2008) from the analysis of a bigger sample of CBe stars (152 stars in 42 open clusters) in open clusters. Later, \cite{2017Arcos} and \cite{2020Gourav} also observed a similar trend in the spectral type distribution for their sample of field CBe stars.

We analyzed the spectral type distribution of a sample of 31 CBe stars, which includes the 13 CBe stars mentioned in this work, 2 probable transient Be stars and an additional sample of 16 CBe stars older than 100 Myr retrieved from the literature. The spectral type distribution plot is shown in Fig. 7, where our sample of CBe stars are shown in red circles. We also included the sample of 152 CBe stars from \cite{2008Mathew} for comparison purpose.

We found that the distribution of our sample is peaking near late B-types. This is expected since we studied older clusters having an age greater than 100 Myr. \cite{2008Mathew} found that their sample of CBe stars in young clusters are mostly peaking at B1-B2 spectral types. This is again expected as they studied young clusters having age lesser than 100 Myr. So now considering the whole cluster age range till 300 Myr, we notice that there actually exist a bimodal distribution in spectral types of CBe stars. Observing such a similar trend in fields and both young and old open clusters is interesting and has to be real. However, we notice another peak emerging at B5 - B6 from the present study. This needs to be understood from the analysis of a large sample of CBe stars, particularly in old open clusters.

\begin{figure}
\centering    
\includegraphics[width=9cm,angle=0]{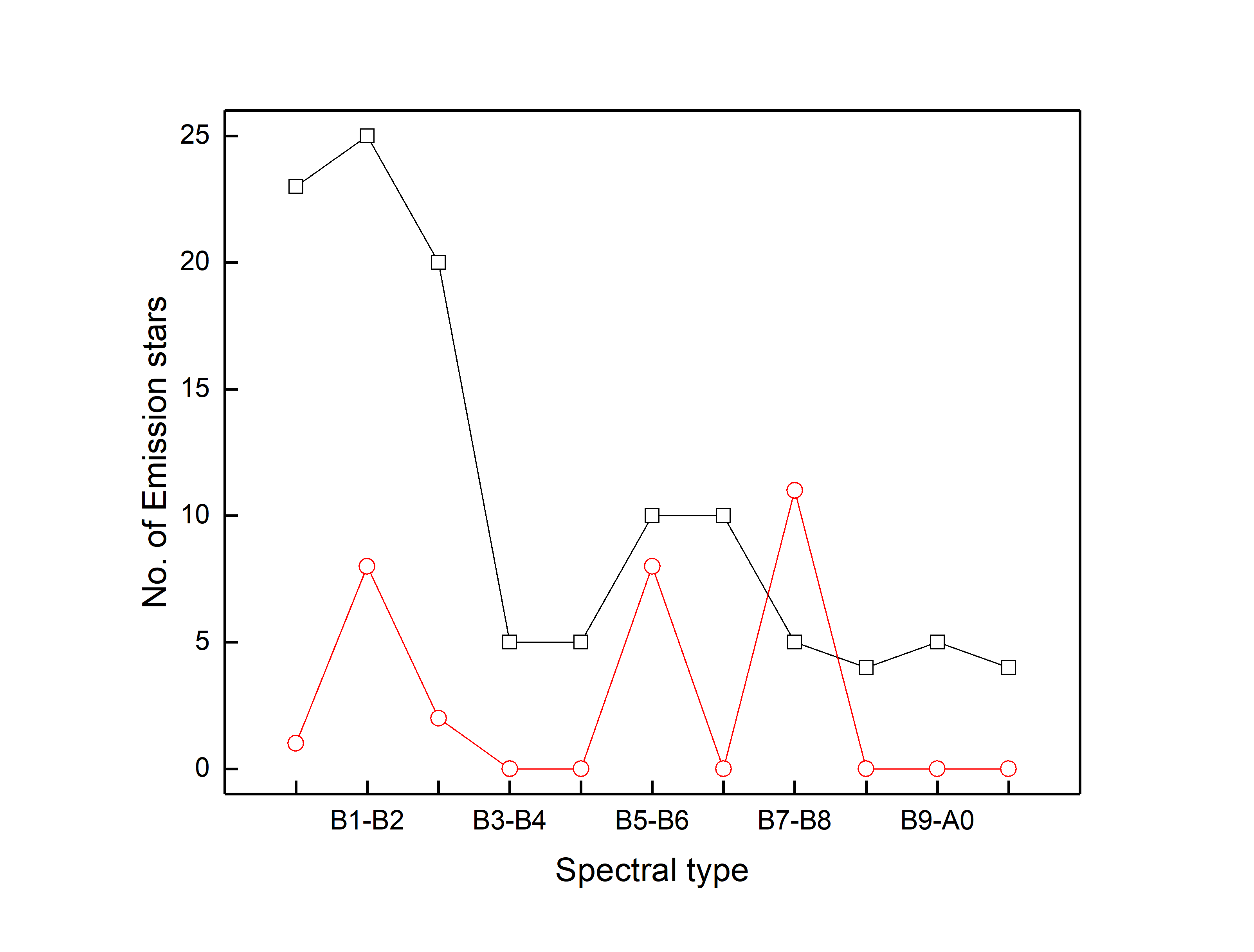} 
\caption{Spectral type distribution of our identified CBe stars in open clusters older than 100 Myr compared to that observed by \cite{2008Mathew}. Here, the red line represents our 31 CBe stars from our survey and literature, whereas the black line represents those CBe stars from young clusters (age below 100 Myr) studied by \cite{2008Mathew}. It is observed that the distribution of our sample stars is also peaking at late B-types as expected.}
\end{figure}

\section{Summary}
In this paper, we performed slitless spectroscopy of 71 open clusters older than 100 Myr to identify and characterize CBe stars in older clusters for the first time. Our present work is a complimentary study with that of \cite{2008Mathew} who studied clusters younger than 100 Myr to identify and characterize CBe stars. The prominent results obtained from our study are summarized below:

\begin{itemize}
\item We identified 13 CBe stars in 11 clusters, one of which (TYC 2679-432-1) is detected as a CBe star for the first time. Moreover, we also found 2 other B-type emission-line stars in 2 separate clusters showing H$\alpha$ in absorption for the first time, as far as literature is concerned. This suggests that both these stars might be passing through disc-less episode presently if they are CBe stars, which can be confirmed by further studies through spectroscopy.
  
\item We estimated the spectral types for all these 15 stars using photometric technique and found that they belong within B0.5 -- B8 types. 

\item We also observed that all our identified 13 CBe stars show both H$\alpha$ in emission and IR excess, which is suggestive that all of them possess gaseous, circumstellar discs.
  
\item Furthemore, our study confirms that a bimodal distribution do indeed exist in CBe star spectral types.

\end{itemize}

Our study will motivate the Be star community about the need to further detect and study CBe stars in other older clusters. Detection and study of more CBe stars in fields and both younger cum older clusters may provide new insights about the `Be pheomenon' in different environments. Recently, \cite{2021Anusha} identified 159 Classical Ae (CAe) stars which are regarded as the lower mass counterparts of CBe stars, thus increasing the total number of detected CAe stars to 180. In a separate study, \cite{2021Li} detected 12 (6 of which are new detections) Oe stars: thought to be the hotter counterpart of CBe stars (\citealt{2013Rivinius}). Hence, further studies of Be stars in different environments can also provide clues about these separate types of emission-line stars too.

\section*{Acknowledgements}
This research has made use of the WEBDA data base, maintained by the Institute for Astronomy of the University of Vienna. Also JHK data from the 2MASS catalog, which is a joint project of the University of Massachusetts and the Infrared Processing and Analysis Centre/California Institute of Technology, funded by the National Aeronautics and Space Administration and the National Science Foundation. 
The following databases are also used:(1) SIMBAD data base, operated at CDS, Strasbourg, France. (2) VOSA, developed under the Spanish Virtual Observatory project supported from the Spanish MINECO through grant AyA2017-84089.(3) VizieR is a joint effort of CDS (Centre de Données astronomiques de Strasbourg) and ESA-ESRIN (Information Systems Division). Special thanks to Mrs. Akshaya Subbanna, PhD scholar and Dr. Sreeja S Kartha, Assistant Professor, Dept. of Physics \& Electronics, CHRIST (Deemed to be University) for all technical discussions.

\begin{table*}
\caption{List of our program clusters and the corresponding log of observations.}\label{Table:1}
\begin{tabular}{llllllll} \hline
Cluster & Date of observation & RA & Dec & Log Age & distance & Exposure Time (s) &      \\
       & (yyyy/mm/dd)           & (hh mm ss)      & (dd mm ss)       &     & (pc)        & R band      & R/Grism 5 \\ \hline
Alessi 53        & 2016-10-23          & 06 29 24      & +09 10 55      & 8.4       & 4570          & 15          & 600           \\
Basel 14        & 2016-06-15          & 21 21 18      & +44 49 00      & 8.3       & 964           & 15          & 600           \\
Basel 4         & 2016-10-11          & 05 48 30      & +30 13 00      & 8.3       & 3000          & 15          & 600           \\
Basel 7         & 2016-10-23          & 06 36 36      & +08 21 00      & 8.033      & 1684          & 15          & 600           \\
Berkeley 15       & 2017-03-22          & 05 02 06      & +44 30 43      & 8.4       & 3300          & 15          & 600           \\
Berkeley 30       & 2015-07-12          & 06 57 42      & +03 13 00      & 8.48       & 4790          & 15          & 600           \\
Berkeley 51       & 2016-09-07          & 20 11 54      & +34 24 06      & 8.18       & 3200          & 15          & 600           \\
Berkeley 61       & 2016-01-02          & 00 48 30      & +67 12 00      & 8.3       & 3335          & 15          & 900           \\
Berkeley 68       & 2016-10-11          & 04 44 30      & +42 04 00      & 8.391      & 1678          & 15          & 600           \\
Berkeley 80       & 2016-10-11          & 18 54 22      & -01 13 00      & 8.48       & 3300          & 15          & 600           \\
Berkeley 84       & 2016-09-07          & 20 04 43      & +33 54 18      & 8.08       & 2025          & 15          & 600           \\
Berkeley 100       & 2015-07-12          & 23 25 58      & +63 46 48      & 8.2       & 3355          & 15          & 600           \\
Berkeley 47       & 2015-08-27          & 19 28 36      & +17 22 06      & 8.2       & 1420          & 15          & 600           \\
Berkeley 49       & 2015-08-27          & 19 59 31      & +34 38 48      & 8.2       & 2035          & 15          & 600           \\
Berkeley 50       & 2016-09-07          & 20 10 24      & +34 58 00      & 8.4       & 2100          & 15          & 600           \\
Berkeley 58       & 2015-07-12          & 00 00 12      & +60 58 00      & 8.4       & 3715          & 15          & 600           \\
Berkeley 6        & 2016-01-01          & 01 51 12      & +61 05 00      & 8        & 2300          & 15          & 600           \\
Berkeley 60       & 2015-07-12          & 00 17 42      & +60 56 00      & 8.2       & 4365          & 15          & 600           \\
Berkeley 90       & 2016-01-01          & 20 35 18      & +46 50 00      & 8        & 2430          & 15          & 600           \\
Berkeley 93       & 2016-01-02          & 21 56 12      & +63 56 00      & 8        & 5600          & 15          & 900           \\
Czernik 19       & 2016-10-11          & 04 57 09      & +28 46 47      & 8        & 2525          & 15          & 600           \\
Czernik 5        & 2016-01-02          & 01 55 43      & +61 21 01      & 8.45       & 2750          & 15          & 900           \\
Ic 2156         & 2015-07-12          & 06 04 51      & +24 09 30      & 8.4       & 2100          & 15          & 600           \\
Kharchenko 1      & 2016-01-01          & 06:08:48      & 24:19:54       &         & No data         & 15          & 900           \\
King 18         & 2016-10-11          & 22 52 06      & +58 17 00      & 8.4       & 2340          & 15          & 600           \\
King 6         & 2015-07-12          & 03 28 06      & +56 27 00      & 8.34       & 871           & 15          & 600           \\
King 15         & 2016-01-01          & 00 32 54      & +61 52 00      & 8.4       & 3162          & 15          & 900           \\
King 20         & 2016-01-01          & 23 33 17      & +58 28 33      & 8.3       & 1900          & 20          & 900           \\
KPS\_2012        & 2016-10-11          & 19:48:32      & 21:57:35       &         & No data         & 15          & 600           \\
MWSC 0817        & 2016-10-11          & 06:14:48      & 12:52:24       &         & No data         & 15          & 600           \\
NGC 0225        & 2016-01-02          & 00 43 39      & +61 46 30      & 8.114      & 657           & 15          & 600           \\
NGC 0744        & 2016-01-02          & 01 58 33      & +55 28 24      & 8.248      & 1207          & 15          & 900           \\
NGC 2215        & 2016-10-23          & 06 20 49      & -07 17 00      & 8.369      & 1293          & 15          & 600           \\
NGC 2251        & 2016-10-23          & 06 34 38      & +08 22 00      & 8.427      & 1329          & 15          & 600           \\
NGC 2254        & 2016-10-23          & 06 35 49      & +07 40 24      & 8.307      & 2364          & 15          & 600           \\
NGC 2269        & 2017-04-18          & 06 43 17      & +04 37 30      & 8.416      & 1687          & 15          & 600           \\
NGC 2323        & 2015-07-12          & 07 02 42      & -08 23 00      & 8.096      & 929           & 15          & 600           \\
NGC 2335        & 2017-04-18          & 07 06 49      & -10 01 42      & 8.21       & 1417          & 15          & 600           \\
NGC 6469        & 2015-08-27          & 17 53 13      & -22 19 11      & 8.36       & 550           & 15          & 600           \\
NGC 6705        & 2016-09-07          & 18 51 05      & -06 16 12      & 8.302      & 1877          & 15          & 900           \\
NGC 7058        & 2016-09-07          & 21 21 53      & +50 49 11      & 8.35       & 400           & 15          & 600           \\
NGC 7067        & 2016-09-07          & 21 24 23      & +48 00 36      & 8        & 3600          & 15          & 600           \\
NGC 7245        & 2016-09-07          & 22 15 11      & +54 20 36      & 8.246      & 2106          & 15          & 600           \\
NGC 7296        & 2016-09-07          & 22 28 02      & +52 19 00      & 8        & 2930          & 15          & 600           \\
NGC 0103         & 2015-07-12          & 00 25 16      & +61 19 24      & 8.126      & 3026          & 15          & 600           \\
NGC 0136         & 2016-01-01          & 00 31 31      & +61 30 36      & 8.4       & 5220          & 15          & 900           \\
NGC 1513         & 2016-01-02          & 04 09 57      & +49 30 54      & 8.11       & 1320          & 15          & 900           \\
NGC 1545         & 2016-10-11          & 04 20 57      & +50 15 12      & 8.448      & 711           & 15          & 600           \\
NGC 1582         & 2016-10-11          & 04 31 53      & +43 49 00      & 8.48       & 1100          & 15          & 600           \\
NGC 1664         & 2016-10-11          & 04 51 06      & +43 40 30      & 8.465      & 1199          & 15          & 600           \\
\hline
\end{tabular}
\end{table*}

\newpage 
\begin{table*}[]
\begin{tabular}{llllllll} \hline

Cluster & Date of observation & RA & Dec & Log Age & distance & Exposure Time (s) &      \\
       & (yyyy/mm/dd)           & (hh mm ss)      & (dd mm ss)       &     & (pc)        & R band      & R/Grism 5 \\ \hline
NGC 1778         & 2016-10-11          & 05 08 04      & +37 01 24      & 8.155      & 1469          & 15          & 600           \\
NGC 6603         & 2016-09-07          & 18 18 26      & -18 24 24      & 8.3       & 3600          & 15          & 600           \\
NGC 6709         & 2015-07-08          & 18:51:18      & 10:19:06       & 8.178      & 1075          & 30          & 600           \\
NGC 6997         & 2016-01-01          & 20 56 40      & +44 38 30      & 8        & 620           & 15          & 600           \\
NGC 7031         & 2016-01-01          & 21 07 12      & +50 52 30      & 8.138      & 900           & 15          & 600           \\
NGC 7062         & 2016-09-07          & 21 23 27      & +46 22 42      & 8.465      & 1480          & 15          & 600           \\
NGC 7086         & 2016-01-01          & 21 30 27      & +51 36 00      & 8.142      & 1298          & 15          & 600           \\
NGC 7226         & 2016-01-01          & 22 10 26      & +55 23 54      & 8.436      & 2616          & 15          & 900           \\
NGC 7762         & 2015-08-27          & 23 50 01      & +68 02 18      & 8.425      & 744           & 15          & 600           \\
NGC 886         & 2016-01-01          & 02 23 27      & +63 46 12      & 8.48       & 1700          & 15          & 900           \\
Platais 1        & 2016-09-07          & 21 30 02      & +48 58 36      & 8.244      & 1268          & 15          & 600           \\
Roslund 1        & 2016-10-11          & 19 45 00      & +17 31 12      & 8.47       & 670           & 15          & 600           \\
Roslund 3        & 2016-10-11          & 19 58 42      & +20 29 00      & 8.036      & 1467          & 15          & 600           \\
Ruprecht 144       & 2015-08-27          & 18 33 34      & -11 25 00      & 8.18       & No data         & 15          & 600           \\
Stock 24        & 2016-01-02          & 00 39 42      & +61 57 00      & 8.08       & 2818          & 15          & 900           \\
Tombaugh 5       & 2016-01-02          & 03 47 48      & +59 03 00      & 8.3       & 1750          & 15          & 900           \\
Trumpler 2       & 2016-01-02          & 02 37 18      & +55 59 00      & 8.169      & 651           & 15          & 900           \\
Trumpler 32       & 2016-09-07          & 18 17 30      & -13 21 00      & 8.48       & 1720          & 15          & 600           \\
Trumpler 34       & 2015-08-27          & 18 39 48      & -08 25 00      & 8.1       & 1738          & 15          & 600           \\
Turner 2         & 2016-06-16          & 18 17 12      & -18 49 30      & 8        & 1190          & 15          & 600           \\
 \hline  
\end{tabular}
\end{table*}

\begin{table*}
\centering
\caption{List of 16 CBe stars found in 6 separate old clusters above 100 Myr as obtained from the literature. Here star mark (*) represents 5 stars whose spectral types were estimated by us. Their corresponding distance (d) and E(B-V) values are taken from the WEBDA database. For the rest 11 cases, spectral types are obtained from the literature.}
\begin{minipage}{130mm}
\begin{tabular}{ccccc}
\hline 
Cluster   & Star &Spectral Type     & Reference & \\\hline
NGC 3114 & NGC 3114 003 & B6 & \cite{2018Aidelman} \\
 & NGC 3114 004 &B8 & \cite{2018Aidelman} \\
 & NGC 3114 028 &B6 &\cite{2018Aidelman} \\
 & NGC 3114 033 &B5 & \cite{2018Aidelman} \\
 & NGC 3114 091 &B8 & \cite{2018Aidelman} \\
 & NGC 3114 129 &B8 & \cite{2018Aidelman} \\ \hline 
NGC 1912 & TYC 2415-940-1 &B8* & \cite{2000H} \\
 & BD+35 1111 &B8* & \cite{2000H} \\
 & TYC 2415-122-1 &B8* & \cite{2000H} \\ \hline 
Pleiades & 23 Tau &B6 & \cite{1982Slettebak})\\
 & eta Tau &B7 & \cite{1982Slettebak}\\
 & 28 Tau &B8 & \cite{2008Taranova}\\ \hline
 
NGC 1857 & GGA 329 &B1.5* & \cite{1990Lasker}\\
 & VES 888 &B1* & \cite{1990Lasker}\\ \hline
NGC 1647 & HD 30123 &B8 & \cite{2000H}\\ \hline
NGC 2301 & HD 50064 &B6 & \cite{1942Merrill} \\ \hline
\end{tabular}
\end{minipage}
\end{table*}

 \begin{table*}
\caption{List of our identified 13 CBe stars in 11 clusters having age above 100 Myr along with their coordinates and V magnitudes as obtained from the SIMBAD database. Here, TYC 2679-432-1 (marked in bold) of the cluster Berkley 50 is the star identified as a CBe star for the first time by us. The same parameters of another 2 stars showing H$\alpha$ in absorption as detected through this survey are also shown in the Table. We estimated the spectral for all these 15 stars using photometric technique.}\label{Table:2}

\begin{tabular}{lllllll}
\hline
Cluster   & Star Name        & RA      & DEC      & V mag & Spectral Type                  \\ 
&&&&& \\ \hline
Tombaugh 5 & SS 16       & 03 47 50 & +59 03 59 & 11.42 & B2 \\
NGC 1778 & HD 280460        & 05 08 13 & +36 59 36 & 9.7  & B3&             \\
    & HD 280461        & 05 08 04 & +36 58 27 & 10.19 & B6           \\
    & HD 280462        & 05 08 02 & +37 03 03 & 10.21 & B6            \\
IC 2156   & LS V +24 11       & 06 05 02 & +24 09 28 & 11.67 & B2                 \\
Ruprecht 144 & SS 398         & 18 33 52 & -11 24 31 & 12.1 & B8                    \\
Trumpler 34 & GSC 5692-0543      & 18 39 49 & -08 25 40 & 12.13 & B6                   \\
NGC 6709   & BD+10 3698       & 18 51 32 & +10 19 10 & 9.65 & B2.5            \\
Berkeley 47 & TYC 1605-346-1     & 19 28 31 & +17 22 22 & 12.34 & B2                  \\
Berkeley 50 & \textbf{TYC 2679-432-1} & 20 10 05 & +34 55 44 & 11.21 & B0.5                   \\
Berkeley 90 & UCAC3 274-184438    & 20 35 41 & +46 46 49 & 12.85 & B2         \\
NGC 7067  & LS III +47 37 & 21 24 12  & +48 00 39 & 13.2 & B1               \\
King 20   & GGR 148       & 23 33 03 & +58 27 44 & 12.11 & B2.5               \\\hline

H$\alpha$ in absorption \\ \hline
NGC 6709 & $[KW97]$ 35-12 & 18 51 10 & +10 23 25 & 10.98 & B8 
 \\
Trumpler 2 & HD 16080 & 02 37 00 & +55 54 41 & 9.32 & B2
  \\ 
 \hline
\end{tabular}

\end{table*}

\label{lastpage}

\end{document}